\documentclass[onecolumn,authoryear]{els-mrw} 
\usepackage{aas_macros}
\usepackage{wrapfig}

\newcommand{\kms}{km\,${\rm s}^{-1}$}
\newcommand{\smy}{$M_\odot\,{\rm yr}^{-1}$}

\usepackage{amsmath,amssymb,amsfonts,amsthm,makeidx,graphicx}
\usepackage{txfonts}
\usepackage{helvet}

\begin{document}

\chapter{Wolf-Rayet stars}\label{chap1}

\author[1]{Tomer Shenar}%

\address[1]{\orgname{Tel Aviv University}, \orgdiv{The School of Physics and Astronomy}, \orgaddress{ Tel Aviv 6997801, Israel}}

\articletag{Chapter Article tagline: update of previous edition,, reprint..}

\maketitle

\begin{glossary}[Glossary]

\term{Radiatively-driven winds} Outflow of material from the stellar surface driven via momentum deposition of the stellar radiation onto the outer layers of the star.

\term{Line-driven winds} Radiatively-driven winds in which the main opacity sources are bound-bound transitions of atoms.

\term{Eddington limit} The maximum luminosity a radiating object can emit before becoming unbound by radiation pressure.

\term{Wolf-Rayet (WR) stars} A spectroscopic class of stars which exhibit strong, broad emission lines belonging to high ionization stages of elements such as He, C, N, and O. Their emission lines originate in powerful, fast, radiatively-driven stellar winds.

\term{Central WR stars of planetary nebulae ([WR] stars)} Hot stripped cores of low- and intermediate-mass stars which are embedded in the ionized ejecta of their progenitors. 

\term{Main sequence WR (MS-WR) stars} Hydrogen-rich, core H-burning WR stars, typically classified WNh. MS-WR stars typically have masses  $\gtrsim 100\,M_\odot$.

\term{Classical WR (cWR) stars} Hot, stripped WR stars that have evolved past core H-burning and are typically core He-burning. Such stars usually have masses in the range $\approx 8 - 30\,M_\odot$. They represent the final evolutionary stages of stars initially more massive than $\approx 20\,M_\odot$ (at solar metallicity) prior to core collapse.

\term{WN stars} Nitrogen-rich WR stars, enriched with N and He via the CNO cycle which occured in the original H-burning core of the progenitor star. WN stars can be either MS-WR stars or cWR stars, though the majority belong to the latter group.

\term{WC/WO stars} Carbon/Oxygen-rich WR stars, showing the main products of He-burning. Given their nature, WC/WO stars are necessarily cWR stars, that is, post main sequence.  WC stars are thought to have originated from WN stars.

\term{Helium stars, stripped stars} Often used synonymously, helium stars / stripped stars are hot, helium-rich stars whose progenitors lost their outer layers via some mechanism. All cWR stars are helium stars, but the converse need not hold, since not all helium stars exhibit the strong winds which define the WR spectral class.

\term{Rosseland mean opacity ($\tau_{\rm Ross}$)}  Weighted average opacity of a medium  which accounts for wavelength dependence and assumes local thermodynamic equilibrium.

\term{Surface effective temperature ($T_*$)} The effective temperature at the surface of the star: the outermost layer at which the star is still in (approximate) hydrostatic equillibrium. The surface effective temperature satisfies the Stephan-Boltzmann equation with respect to the stellar radius $R_*$ and luminosity $L$ and the surface (i.e., $L \propto R_*^2 T_*^4 $).

\term{Photospheric effective temperature ($T_{2/3}$)} The effective temperature at the radius in which the majority of stellar light is escaping from, situated roughly at a radial layer at which the Rosseland optical depth reaches $\tau_{\rm Ross} = 2/3$. Due to their optically-thick winds, The photospheres of WR stars lie above their stellar surfaces, and their photospheric effective temperatures are typically lower ($T_{2/3} < T_*$).

\term{The Conti scenario / the single-star channel} A formation channel of cWR stars which explains their stripping via intrinsic mass-loss of their progenitor stars (continuous or eruptive mass loss). 

\term{The binary channel} A formation channel of cWR stars which explains their stripping via binary interactions, typically mass transfer. 


\end{glossary}

\begin{glossary}[Nomenclature]
\begin{tabular}{@{}lp{34pc}@{}}
BH & black hole \\
CE & common envelope \\
cWR & classical Wolf-Rayet\\
GW & gravitational wave \\
LGRB & long-duration $\gamma$-ray burst\\
LMC & Large Magellanic Cloud\\
LTE & local Thermodynamic Equilibrium \\
MS-WR & Main sequence Wolf-Rayet\\
SMC & Small Magellanic Cloud\\
SN & Supernova\\
WN/WC/WO & Wolf-Rayet star of the N/C/O-sequence\\
WNE & early-type WN\\
WNL & late-type WN\\
WR & Wolf-Rayet\\
\end{tabular}
\end{glossary}

\begin{abstract}[Abstract]
Massive Wolf-Rayet (WR) stars comprise a spectroscopic class characterized by high temperatures ($T_{\rm eff} \gtrsim 30\,$kK) and powerful and rapid stellar winds. Hydrogen-rich WR stars represent the most massive stars in existence ($M \gtrsim 100\,M_\odot$), while classical WR stars are hydrogen-depleted, evolved massive stars which probe the final evolutionary stages of massive stars prior to core collapse. They dominate entire stellar populations in terms of radiative and mechanical feedback, and are thought to give rise to powerful transients such as hydrogen-stripped supernovae (type Ibc SNe) and long-duration $\gamma$-ray bursts (LGRBs). In this chapter, we summarize the main observed properties of WR populations in our Galaxy and nearby galaxies, and discuss open problems in our understanding of their structure and formation.
\end{abstract}

\begin{BoxTypeA}[]{Key points}
\begin{itemize}
\item Wolf-Rayet (WR) stars comprise a rare spectroscopic class of stars with broad, strong emission lines belonging most prominently to He\,{\sc ii}. These lines originate in powerful radial outflows driven by the proximity of the stars to the Eddington limit. There are close to 700 WR stars known in our Galaxy out of $\approx 2000$ estimated in total; thousands of others are known outside our Galaxy. WR stars originate from stars initially more massive than $M_{\rm ini}\gtrsim 20\,M_\odot$ (O-type stars) in the Milky Way, and from higher initial masses at lower metallcities.

\item WR stars can easily dominate the radiative and mechanical feedback of entire stellar populations. They are thought to be progenitors of violent explosions such as type Ibc supernovae and potential direct progenitors of black holes (BHs), and are sources of heavy elements and dust.  BH+BH mergers detected via gravitational-wave (GW) events likely pass a WR binary phase during their evolution. 

\item Massive WR stars come in two main types: {\bf main sequence WR stars} (MS-WR) and {\bf classical WR star} (cWR). MS-WR stars represent very massive ($M_{\rm ini} \gtrsim 100\,M_\odot$) core H-burning stars, while cWR stars are hydrogen-depleted objects that have evolved off the main sequence and suffered intense mass loss.  The process by which the progenitors of cWR stars lost their envelopes involves either intrinsic mass loss (the single-star channel) or binary interactions (the binary channel). The observed binary fraction of cWR stars is $\approx 40-50\%$, and appears to be metallicity independent.

\item Two key problems are still considered unsolved in the field. First, semi-analytic and numeric hydrodynamic calculations of WR winds reveal the presence of strong instabilities, non-radial outflows, inflated surfaces, and non-analytical wind velocity fields. These phenomena challenge traditional 1D approaches for deriving the stellar and wind parameters of WR stars. Secondly, the formation routes of cWR stars are still uncertain. The abundance of WR stars without apparent companions suggests that a significant fraction might have formed as single stars, but contemporary evolution models struggle to form them.

\end{itemize}
\end{BoxTypeA}

\section{Introduction}\label{sec:WRIntro}

\subsection{Definition and observables}\label{subsec:Def}



\begin{figure}[!h]
\centering
\includegraphics[width=\textwidth]{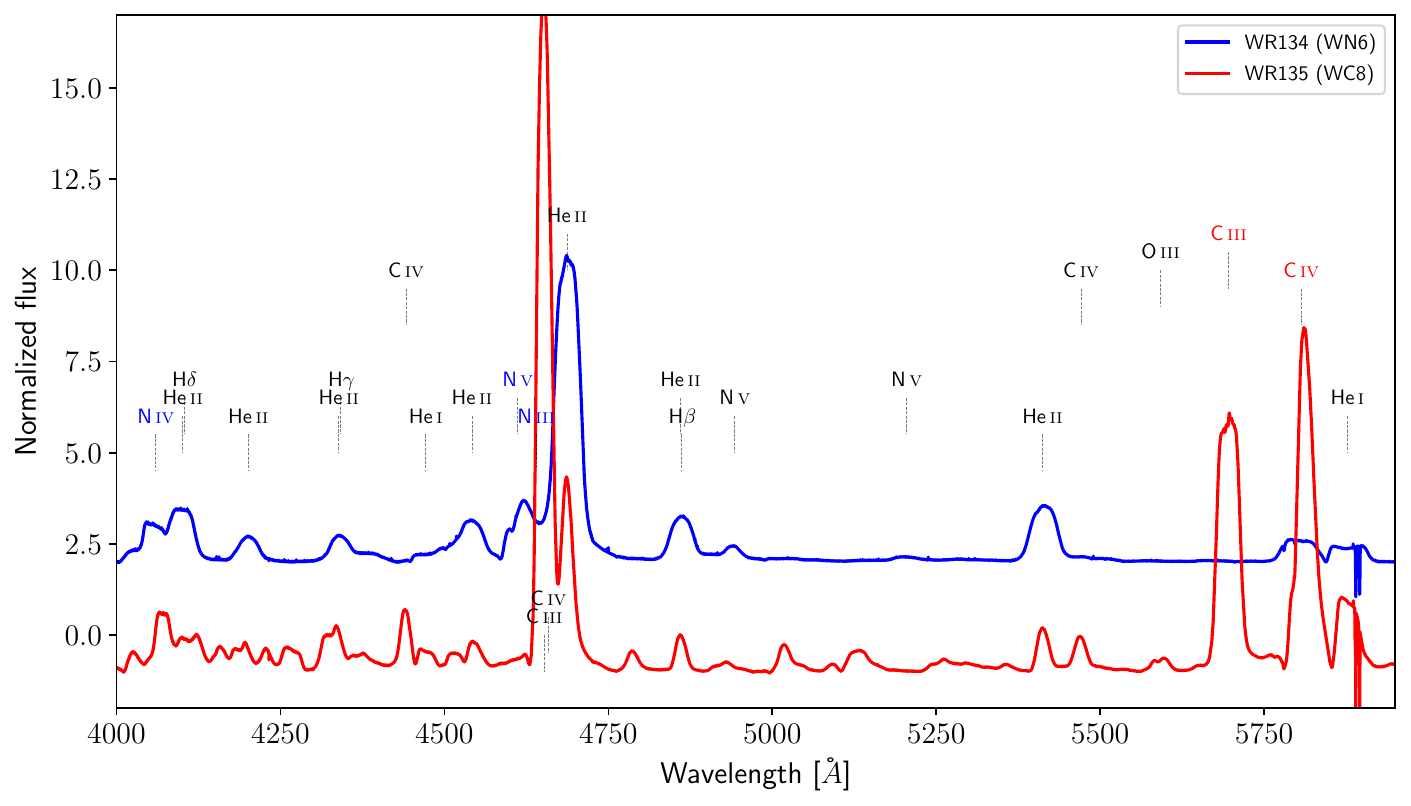}
\caption{Spectra of WR\,134 (blue) and WR\,135 (red) obtained with the HERMES spectrograph on the 1.2m Mercator telescope in La Palma, Canary Islands, Spain. WR\,134 and 135 are some of the first WR stars discovered by \citet{WolfRayet1867}, belonging to the WN and WC classes, respectively. The spectra have a resolving power of $R\approx 80,000$ and a signal-to-noise ratio (SNR) of $\approx 300$. Important spectral lines are marked, with diagnostic lines for the respective WN and WC spectral classes in blue and red color, respectively. The spectrum of WR\,135 is shifted for clarity. Adapted from \citet{Dsilva2020,Dsilva2022}. }
\label{fig:WRSpec}
\end{figure}

Over a century and a half ago, astronomers Charles Wolf and Georges Rayet discovered a unique set of stars which exhibited peculiar optical spectra (Fig.\,\ref{fig:WRSpec}): unlike the usual absorption-line dominated spectra of stars, the spectra of their newly discovered objects  showed broad (width of a few $1000$ \kms), strong (equivalent widths of tens of $\AA$) emission lines associated with high ionization stages of atoms such as helium, nitrogen, and carbon \citep{WolfRayet1867, Crowther2007}. Stars which exhibit such spectra are classified as WR stars, regardless of their physical nature. It has been long established that these features generally stem from powerful, dense, radiatively-driven stellar winds \citep{Beals1929}\footnote{In the WR-like star HD\,45166, a powerful magnetic field is the main cause for strong emission lines  \citep{Shenar2023}; such cases are likely extremely rare.}. The main opacity sources driving these winds are millions of line transitions belonging to heavy elements \citep{Lucy1993}. 
Powerful stellar winds are launched when the stellar luminosity approaches the Eddington limit: the critical luminosity at which the star is no longer bound against radiation pressure for a fully ionized hydrogen atmosphere (electron scattering):

\begin{equation}
    L_{\rm Edd} = \frac{4 \pi G M m_{\rm p} c}{\sigma_{\rm e}},
\end{equation}
where $G$ is the gravitational constant, $m_{\rm p}$ is the proton mass, $c$ is the speed of light, $\sigma_{\rm e}$ is the Thompson electron scattering opacity, and $M$ is the stellar mass. One defines $\Gamma = L/L_{\rm Edd}$ as the Eddington factor for a star with luminosity $L$. For WR stars, the Eddington factor reaches typical values of the order of $\Gamma \approx 0.5$. It is important to note that the Eddington limit is a proxy for the strength of radiation pressure, but the main source of opacity lies in line transitions rather than electron scattering \citep{Graefener2011, Sander2020}. the Eddington factor scales as $L/M$, implying that stars tend to become WR when their luminosities are very high compared to their masses.

The presence of highly ionized elements in the atmosphere of WR stars implies that they are hot, with deduced surface effective temperatures in the range $T_* \approx 20 - 150\,$kK \citep{Hamann2006, Sander2020}. Massive WR stars, which are the focus of this chapter (see below), span luminosities in the range $L \approx 10^{5}$ - $10^{7}\,L_\odot$, and lose mass at rates in the range $\dot{M} \approx 10^{-5} - 10^{-4}\,$\smy. With lifetimes of $0.5 - 1\,$Myr, this implies that WR stars may lose tens of solar masses worth of mass throughout their evolution!

WR spectra are typically seen in two main flavors: those which, aside from helium, show lines belonging mainly to nitrogen (typically N\,{\sc iii, iv}, and {\sc v}) -- the WN class -- and those which show mainly lines belonging to carbon (typically c\,{\sc iii} and {\sc iv}) -- the WC class (for details on WR classification, see \citealt{Smith1996}). Rarely, oxygen-rich spectra are observed, giving rise to the rare WO class. Similarly rare are transition classes which show features characteristic of WN and WC stars, denoted WN/WC stars \citep{Massey1989}. The WN and WC classes are further divided to subclasses types WN2 -- WN11 and WC4 -- WC9 based on the ratios of nitrogen and carbon lines for WN and WC stars, respectively, corresponding to their photospheric temperatures (Sect.\,\ref{sec:winds}). One often refers to WN stars in the range WN2 -- 5 as early-type WN stars or WNE, and WN6 -- WN11 as late type WN stars or WNL.

The WR phenomenon occurs both in the low and intermediate-mass (initial masses $M_{\rm ini} \lesssim 8\,M_\odot$) and high-mass regimes. While the nature of low-mass and high-mass WR stars is different, the underlying cause for the WR phenomenon is the same: proximity to the Eddington limit \citep[see e.g.,][]{Toala2024}. Low-mass stars may reach the WR phase after ejecting their outer layers in the planetary nebula phase and exposing their hot cores prior to the white-dwarf phase \citep{Schoenberner1981, Tylenda1993, Crowther1998}. To distinguish them from their massive counterparts, central WR stars of planetary nebulae are denoted as [WR]. Such WR stars are almost always classified as [WC], though a few rare cases of [WN] stars exist \citep{Miszalski2012, Todt2013}. The remainder of the chapter focuses on massive WR stars, with the occasional omission of the explicit term 'massive'.

The blending of the hydrogen Balmer lines with the helium Pickering lines  makes the measurement of the hydrogen content in the atmospheres of WR stars challenging, but the changing strength of blended lines and pure Pickering lines ("the Pickering decrement") allows for an estimate \citep{Smith1996}. Some WR stars exhibit a near-solar content of hydrogen ($X_{\rm H} \approx 60-70\%$ in mass fraction), but the majority ($\gtrsim 90\%$) of WR stars are either partially or  fully stripped of hydrogen \citep{Schmutz1989, Hamann2006}. Only WN stars were found to show hydrogen in their atmospheres, and these are often classified as WNh or WN(h) stars. 

Masses of WR stars are most robustly obtained via analyses of WR binaries, ideally by combining spectroscopy with photometry, interferomtry, or spectropolarimetry  \citep{Lamontagne1996, Massey2002, Richardson2021, Shenar2021}. Otherwise, the mass needs to be estimated from the luminosity via mass-luminosity relations \citep{Graefener2011, Hamann2019, Sander2019}. Typical masses of WR stars range between $\approx 8\,M_\odot$ up to $\approx 200\,M_\odot$, depending on the type of WR stars (see Sect.\,\ref{subsec:EvStat}).

WR stars bear an enormous impact on their surroundings via their radiative and mechanical energy. Even though their numbers are much smaller than their O-star progenitors, the radiative and mechanical energy of WR stars is comparable to that of entire OB-type populations \citep{Doran2013, Ramachandran2018}. Moreover, the high effective temperatures reached by some WR stars make them exclusive sources for He\,{\sc ii} ionizing photons, though lower-mass helium stars may contribute here significantly as well (\citealt{Goetberg2018}; see Sect.\,\ref{subsec:HeStars}). WC specifically are also known to be important dust producers \citep{Williams1987, Usov1991}, a process which was recently seen in action with JWST in the WC binary WR\,140 \citep{Lau2022}.  Finally, owing to their high initial masses, WR star comprise excellent candidate BH progenitors. Hence, WR binaries represent a key phase through which BH+BH mergers detected with the LIGO/VIRGO/KAGRA gravitational-wave (GW) detectors have likely gone through \citep[e.g.][]{Belczynski2010}.


\subsection{How are WR stars found?}\label{subsec:found}

The most prominent spectral feature of WR stars almost regardless of the spectral subtype is the He\,{\sc ii}\,$\lambda 4686$ line, which is often blended with other strong lines (e.g., belonging to carbon; see Fig.\,\ref{fig:WRSpec}). Hence, the most common way of finding WR stars in our Galaxy and other galaxies in via narrow-band imaging surveys which include the He\,{\sc ii}\,$\lambda 4686$ complex. By subtracting the surrounding continuum flux from the flux of stars in this range, one can identify WR populations. A spectroscopic follow-up then leads to their robust classification \citep[e.g.,][]{Neugent2019}. While the He\,{\sc ii}\,$\lambda 4686$ region is a common choice, other emission features can be utilized, for example when only the infrared is accessible due to strong interstellar extinction.

\begin{wrapfigure}{r}{.56\textwidth} 
\vspace{-20pt}
\centering
\includegraphics[width=0.56\textwidth]{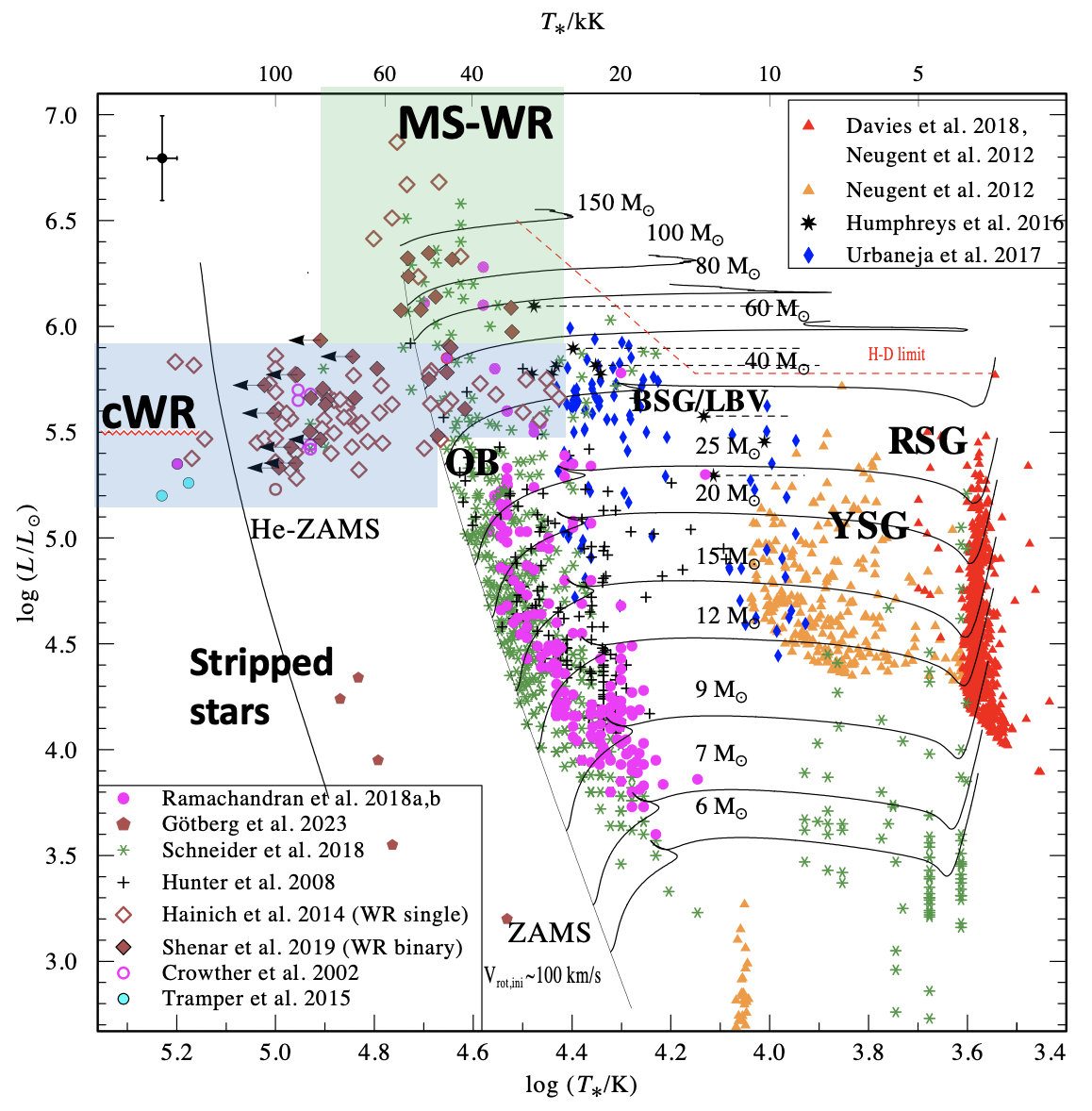}
\caption{HRD of stellar populations in the LMC, adapted from \citet{Ramachandran2019} and \citet{Gilkis2021}. The parameters are retrieved from \citet{Hainich2014,Hainich2015,Shenar2016,Shenar2019,Crowther2002,Tramper2015} for WR stars, \citet{Goetberg2023} for stripped stars,  \citet{Neugent2010,Neugent2012} for Yellow Supergiants (YSGs), \citet{Davies2018,Neugent2012} for Red Supergiants (RSGs), \citet{humphreys_social_2016,kalari_how_2018} for Luminous Blue Variables (LBVs), \citet{trundle_understanding_2004,trundle_understanding_2005,urbaneja_lmc_2017} for Blue Supergiants (BSGs), and \citet{Ramachandran2018,schneider_vltflames_2018,hunter_vlt_2008,castro_spectroscopic_2018,dufton_census_2019,bouret_massive_2013} for populations of OB stars. Evolutionary tracks (black solid lines), accounting for rotation with $\varv_{\mathrm{rot, init}} \,\approx$\,100 km\,s$^{-1} $ for the LMC \citep{Brott2011,Koehler2015}, are labeled.
The rough locations of MS-WR stars and cWR stars and the Humphreys-Davidson (H-D) limit are also marked.}
    \label{fig:HRDLMC}
  \vspace{-30pt}
\end{wrapfigure}  


\subsection{Evolutionary status: main-sequence WR stars and classical WR stars}\label{subsec:EvStat}

Massive WR stars are separated into two main evolutionary categories, illustrated on a Hertzsprung-Russell Diagram (HRD) of the Large Magellanic Cloud (LMC) in Fig.\,\ref{fig:HRDLMC}:

\vspace{0.2cm}

\begin{itemize}  \setlength{\itemsep}{5pt}

    \item  \textbf{\emph{Main-sequence WR (MS-WR) stars}} are hydrogen-rich WR stars (typically classified WNh) that are core hydrogen-burning. The WR phenomenon occurs due to the immense luminosities of the most massive stars, driving $L/M$ to high values. Stars with initial masses $\gtrsim 80-100\,M_\odot$ therefore appear as WR stars already on the main sequence \citep{deKoter1997}. Roughly 10\% of the known massive WR stars are thought to belong to this class. 

\item \textbf{\emph{Classical WR (cWR) stars}} already exhausted hydrogen in their cores and are therefore post main sequence, typically core He-burning. Roughly 90\% of the known WR stars are thought to belong to this category. Unlike MS-WR stars, the WR phenomenon in cWR stars is driven by previous mass loss, that is, their relatively low masses push them closer to the Eddington limit. It is important to note that the presence of hydrogen on the atmospheres of WR stars does not imply that they are core H-burning. For example, all WN stars in the Small Magellanic Cloud (SMC) exhibit hydrogen, but their surface temperatures leave no doubt that they are evolved objects \citep{Hainich2015, Shenar2016, Schootemeijer2018}. Hence, in spite of what is commonly stated in the literature, cWR stars are not necessarily H-free. 

\end{itemize}

\vspace{0.2cm}

MS-WR stars probe the most massive stars known. Analyses of MS-WR binaries indicate masses in the range 100-150\,$M_\odot$ for most MS-WR stars \citep{Tramper2016, Barba2022, Tehrani2019, Bestenlehner2022, Shenar2017WR, Shenar2021}. The most massive stars currently known is RMC\,136 a1 (alias R\,136 a1), with an estimated mass in the range $200-300\,M_\odot$ \citep{Bestenlehner2020, Brands2022}, albeit relying on $L-M$ calibrations given that no companions to it were thus far detected  \citep{Shenar2023_A1}.

In contrast,  cWR stars show clear indications that their progenitor stars ejected their outer, H-rich layers.  They typically weigh $\approx 8 - 20\,M_\odot$ in the Milky Way galaxy \citep{VanDerHucht2001} and reach higher masses at lower metallicities \citep{Hainich2015, Shenar2016}. cWR stars are thought to evolve from massive O-type stars to WN stars and then to WC/WO stars, exposing gradually the products of the CNO cycle and later He burning via intense mass-loss. 
The formation process of cWR stars, that is, the process by which their progenitors lost their outer layers, is thought to occur via two main channels: intrinsic mass-loss or binary mass-transfer. This will be discussed in more detail in Sect.\,\ref{sec:formation}.

\subsection{The relationship between Wolf-Rayet stars and helium stars} \label{subsec:HeStars}

\begin{wrapfigure}{r}{0.5\textwidth} 
\vspace{-10pt}
\centering
\includegraphics[width=0.5\textwidth]{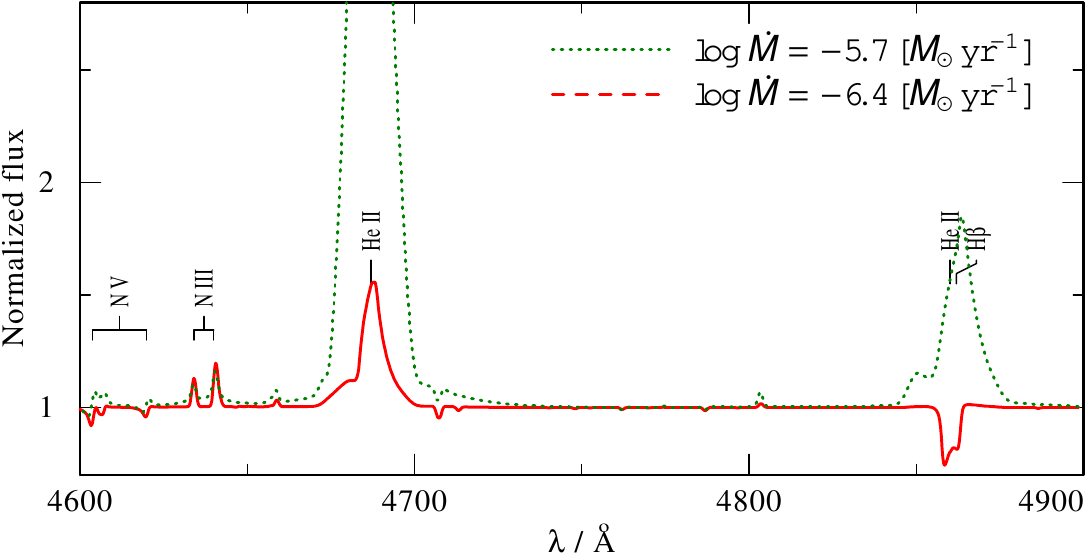}
\caption{Two PoWR models computed with fixed stellar parameters ($\log L/L_\odot = 5.3$, $T_* = 50\,$kK), but two different mass-loss rates (see legend). One can see that by lowering the mass-loss rate, the emission features weaken, causing the spectral appearance to transition from that of a WR star (green) to that of a non-WR helium star (red). Adapted from \citet{Shenar2020}. }
    \label{fig:Rt}
\end{wrapfigure}

Helium stars, also termed stripped stars in the literature, are hot, helium-rich stars which lost their outer layers by some mechanism. Such objects span the entire mass regime. All cWR stars are helium stars, but the converse does not hold, since being a helium star does not necessarily imply a WR-like spectrum.  Low-mass helium stars are termed OB-type subdwarfs (sdB, sdO; \citealt{Heber2009, Geier2019}), rarely observed also as helium giants  or "extreme helium stars" \citep{Jeffery1992, Laplace2020, Gilkis2023}. At the upper-mass end, helium stars take the form of cWR stars. Recently, helium stars in an intermediate mass regime $1\lesssim M \lesssim 8\,M_\odot$ were discovered in the Magellanic Clouds \citep{Drout2023, Goetberg2023}. What distinguishes WR stars from general helium stars is the presence of winds that are strong enough to give rise to their characteristic emission-line spectra (Fig.\,\ref{fig:Rt}).

\begin{figure}[t]
\centering
\includegraphics[width=\textwidth]{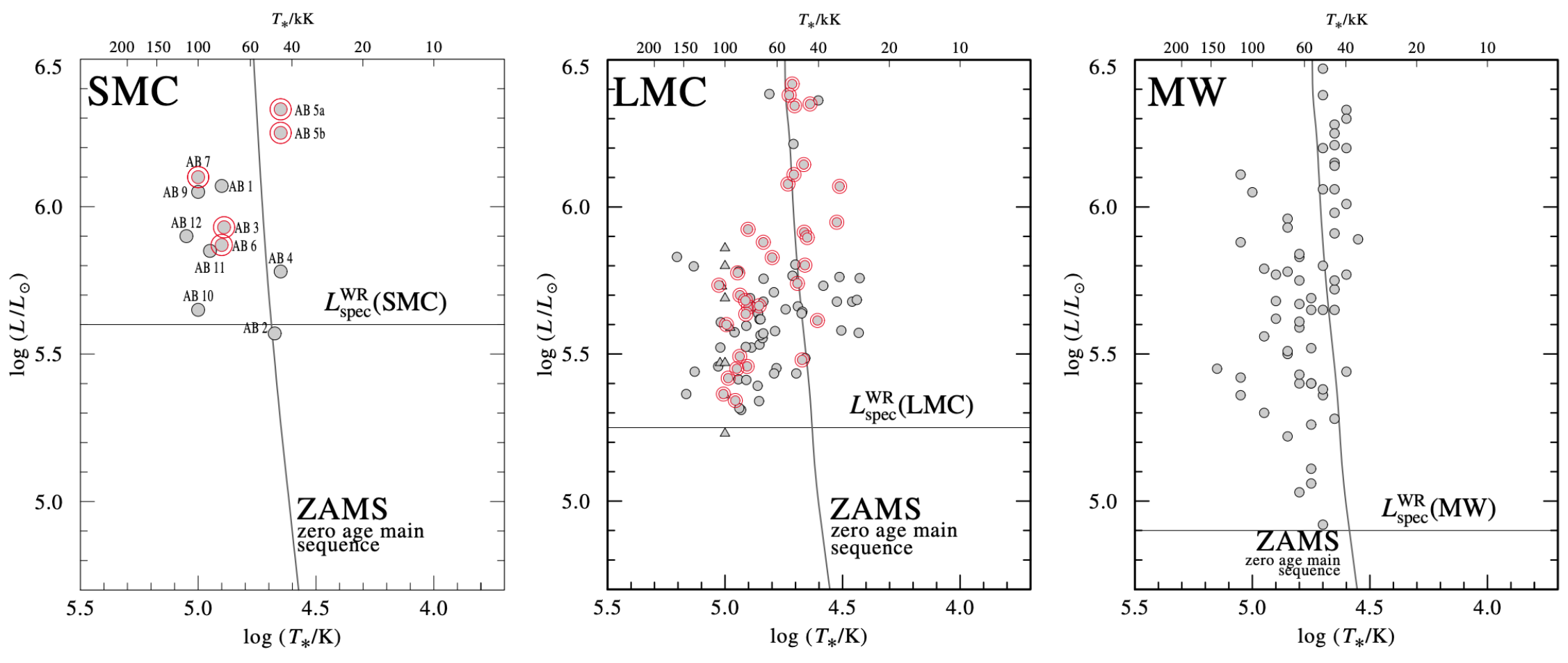}
\caption{HRD positions of SMC, LMC, and MW WN single-star (gray symbols) and binary (gray symbols with red circles) populations for the SMC (left), LMC (middle), and MW (right), adapted from \citet{Shenar2020}. The effective temperatures and luminosities are taken from \citet{Hainich2014, Hainich2015} and \citet{Shenar2016, Shenar2017WR, Shenar2018, Shenar2019}. The minimum luminosities at which the WR phenomenon switches on are marked with horizontal lines ($L^{\rm WR}_{\rm spec}$), showing the increasing trend with decreasing metallicity. The least luminous WN stars in the MW had initial masses of $\approx 20\,M_\odot$, increasing to $\approx 35-40\,M_\odot$ for the SMC \citep{Shenar2020}.  }
\label{fig:HRD}
\end{figure}

As the mass of the stripped helium star decreases, the luminosity drops more sharply, implying that their $L/M$ ratio decreases with mass. One may therefore expect a threshold luminosity below which the WR phenomenon switches off \citep{Dionne2006, Goetberg2018, Sander2019, Shenar2019, Shenar2020}.  Moreover, since the wind strength of WR stars is $Z$-dependent (Sect.\,\ref{sec:winds}), one may expect this threshold to depend on $Z$ too. Indeed, this is precisely what is observed, as is illustrated in Fig.\,\ref{fig:HRD}.

In the Galaxy, at solar metallicity, only helium stars whose progenitors were more massive than about $20\,M_\odot$ (i.e., with birth masses as helium stars of $\approx 8\,M_\odot$) are close  enough to the Eddington limit to  exhibit WR-like spectra. Since the winds of WR stars are weaker at low $Z$ (Sect.\,\ref{sec:winds}), this threshold increases too at low $Z$. \citet{Shenar2020} estimated that the WR phenomenon will only occur for helium stars initially more massive than $\approx 25\,M_\odot$ in the LMC and $\approx 37\,M_\odot$ in the SMC (see also \citealt{Pauli2023}). The WR phenomenon therefore becomes inherently rare at low $Z$, regardless of the formation channel of WR stars (Sect.\,\ref{sec:formation}).

\subsection{Binary frequency}\label{subsec:bin}

As will be discussed in Sect.\,\ref{sec:formation}, binary interactions offer a viable, perhaps dominant, formation route of cWR stars \citep{Paczynski1967}. Binaries therefore not only enable an accurate measurement of WR masses, which are otherwise difficult to obtain (Sect.\,\ref{sec:WRIntro}), but are also a crucial testbed for evolution models of cWR stars. 

The observed binary fraction of WR stars in the Milky Way (MW) galaxy has been reported to be $\approx 40\%$ in earlier studies \citep{VanDerHucht2001}, though this has not been properly bias-corrected until recently. \cite{Dsilva2020, Dsilva2022, Dsilva2023} performed a modern high-resolution spectroscopic survey of a sample of 39 WR stars (12 WC and 27 WN) and derived an observed binary fraction of $f_{\rm bin, obs} ({\rm WC})= 0.58\pm0.14$ for WC stars and $f_{\rm bin, obs} ({\rm WN}) = 0.41 \pm 0.09$ for WN stars, amounting to a total binary fraction of $f_{\rm bin, obs} ({\rm WR}) = 0.46\pm0.11$. The bias correction for these samples is uncertain since the underlying mass and period distributions could greatly differ from those of their natal populations, but the latter studies reported $f_{\rm WC, int} > 0.74$ and $f_{\rm WN, int} = 0.52 \pm 0.13$ for the intrinsic binary fractions of the Galactic WC and WN populations, respectively.

As will be discussed in  Sect.\,\ref{sec:formation}, it has been predicted that the binary fraction of WR stars should increase with decreasing metallicity ($Z$).
The SMC and LMC, with metallicity contents of $\approx 1/5\,Z_\odot$ and $\approx 1/2\,Z_\odot$, respectively \citep{trundle_understanding_2004, hunter_vlt_2008}, offer an excellent probe of metallicity effects on WR populations, including binary properties. \citet{Bartzakos2001}, \citet{Foellmi2003LMC, Foellmi2003SMC} and \citet{Schnurr2008} conducted a spectroscopic radial-velocity (RV) monitoring of the WC and WN populations of the LMC and SMC, and reported binary fractions of $\approx 30-40\%$ in these galaxies, compatible with the observed binary fraction in the MW. Recently, \citet{Schootemeijer2024} conducted a high-resolution RV monitoring of the 7/12 apparently-single WR stars in the SMC. The latter survey had a high sensitivity to periods up to few years and mass ratios as low as 0.1, but did not yield any new binary detections. It thus appears that the majority of WR stars in the SMC truly do not harbor any companions in the parameter regime relevant for binary interactions. There are so far no indications that the binary fraction of WR stars is $Z$-dependent. Why that might be, and what this may imply, will be discussed in Sect.\,\ref{sec:formation}.

\section{Galactic and extragalactic Wolf-Rayet populations} \label{sec:Census}

An updated Galactic WR catalogue \citep{Rosslowe2015}, which is an extension of the 7$^{\rm th}$ WR catalogue by \citet{VanDerHucht2001}, is maintained by P.\ Crowther and is accessible online\footnote{https://pacrowther.staff.shef.ac.uk/WRcat/Description.php}. It currently lists 679 WR stars: 397 WN stars, 278 WC stars, and 4 WO stars. The slight dominance of WN stars could be related to evolutionary timescale arguments, although the sample is thought to be largely incomplete, with a total estimated number of $\approx 2000$ WR stars in our Galaxy. Generally, WC stars are fainter, which could make them more susceptible to observational bias \citep{Crowther2007}. Moreover, roughly 1/3 of the WN stars are classified WNh or WN(h), a subset of which could be MS-WR stars; when comparing H-poor or free WN stars with WC stars, their numbers become comparable.

The SMC and LMC offer the nearest extragalactic populations of WR stars, and have the advantage of being likely complete and probing subsolar-metallicity conditions \citep{Neugent2018}. The LMC harbors a population of 154 WR stars: 126 of which classified as WN, and 28 as WC. The SMC hosts only 12 WR stars, 11 of which are WN, and, peculiarly, one WO star. The sharp drop in the WC/WN ratio is almost certainly related to the drop in metallicity; in such conditions, the stellar winds of WR stars are weaker and prevent them from exposing their C-rich layers, inhibiting the formation of WC stars.

Additional well-studied populations of WR stars include those in the galaxies M31 and M33 \citep{Neugent2011, Neugent2012, Neugent2019}, which host 154 and 206 known WR stars, respectively. The WN/WC ratios in these galaxies is more similar to that of the Milky Way, though an increasing trend is observed towards the outskirts of M33, thought to be related to the metallicity gradient of the galaxy. More distant WR populations are also known, but can only be studied in unresolved populations with current instrumentation \citep{Brown2002, Hadfield2006, Kehrig2013, Brinchmann2008}. For example, the Antenna galaxy, which is a merger of two galaxies, is thought to harbor roughly 4000 WR stars via studies of integrated light \citep{GomezGonzalez2021}.

\section{Wolf-Rayet winds}\label{sec:winds}

Quantifying the strength of WR winds is crucial for correctly predicting their future evolution. WR stars can lose more than half their mass during their lifetimes, affecting the final product type (neutron star vs.\ BH), its mass, and the properties of the accompanying supernova explosion \citep{Dessart2011, Yoon2017, Gilkis2022}. When in binaries, these mass-loss rates also cause an evolution of the orbital periods due to mass and angular-momentum loss, with higher mass-loss rates leading to longer periods. Such changes in the final mass and periods can crucially impact the rates of BH+BH mergers observed with LIGO/VIRGO/KAGRA \citep[e.g.,][]{Belczynski2010}.

Early attempts to model the wind driving of WR stars via radiative processes revealed that the main source of opacity responsible for the efficient deposition of photon momentum onto the wind are millions of bound-bound transitions of atoms in the outer layers of the star \citep{Castor1970, Lucy1993}. These "line transitions" belong primarily to heavy elements, and are typically dominated by the iron-group elements. Early analytical solutions by \cite*{Castor1975}, now known as the CAK formalism, suggested that the velocity field of massive-star winds, and specifically WR winds, can be approximated via a  $\beta$-law: $\varv(r) \propto (1 - \left(R_*/r\right))^\beta$, where $\beta$ is a number of the order of unity and $R_*$ is the radius of the outermost hydrostatic layer. The opacity is further enhanced via Doppler shifts of the expanding material, allowing it to absorb photons along a wider frequency range ("line deshadowing"; \citealt{Owocki1988, Feldmeier1995, Sundqvist2013, Driessen2019}).

The intense radiation fields and relative low densities in WR winds imply that they strongly deviate from local thermodynamic equilibrium (LTE); consistent modelling of WR winds therefore needs to relax LTE (non-LTE). Several codes exist which treat this problem in expanding atmospheres, such as CMFGEN \citep{Hillier1998} and PoWR \citep{Graefener2002, Hamann2003}.  1D wind models utilize either Monte-Carlo approaches \citep[e.g.][]{Lucy1970} or semi-analytical computations of the radiation field \citep{Graefener2005, Sander2017, Sander2020}. Such modern calculations suggest that the velocity fields of WR stars can significantly deviate from a standard $\beta$-law, especially in the inner parts of the wind. While the latter studies are limited to monotonically increasing fields, analytical solutions suggest that the velocity fields may depart from monotonicity in the inner parts, mimicking a "dynamical inflation" of the star \citep{Graefener2012, Sanyal2015, Poniatowski2021}. This inflation could provide an explanation for the inflated radii deduced for WR stars compared to theoretical expectation \citep{Lefever2023}.

Mass-loss prescriptions of WR stars are crucial components of stellar-evolution models of massive stars. Such prescriptions have mostly been retrieved via power-law fits to empirical mass-loss measurements \citep[e.g.][]{Nugis2000, Hainich2015, Shenar2020Corr}, though recently, theoretical prescriptions based on consistent 1D hydrodynamical modelling became available \citep{Sander2020}. 
Both theory as well as observations show that the strength of the winds of WR stars decreases with decreasing metallicity \citep{Vink2005, Bestenlehner2014, Hainich2015, Shenar2019}. Since metals are the main opacity source driving the winds, a lower metal content implies weaker winds. Empirical correlations in the Galaxy, LMC, and SMC, suggest a close-to linear relationship, $\dot{M}_{\rm WR} \propto Z$. A similar trend is found from theoretical calculations \citep{Sander2019}.

The winds of WR stars are clumped, as implied from time-dependent spectroscopic studies and consistent multiwavelength analyses \citep{Hillier1984, Hamann2006, Oskinova2007, Lepine1999}. These clumps are generally thought to arise from instabilities in the wind driving \citep{Owocki1988, Feldmeier1995}. Such inhomogeneities lead to a deviation from 1D radial symmetry, which can be addressed in the framework of 2D or 3D modelling efforts. First 3D models recently been presented by \citet{Moens2022}, albeit approximating the full non-LTE radiative transfer to avoid extensive computation times. The average velocity fields implied from these studies retrieve a law closer to a $\beta$-law. It therefore still remains unclear whether simple analytical formulae can adequately describe the winds of WR stars.

\section{Formation and evolution of cWR stars}\label{sec:formation}

\subsection{The single and binary star channel} \label{subsec:sinbin}

One of the biggest unsolved problems in the field of WR stars concerns the formation of cWR stars: how did their progenitors lose their outer, H-rich envelopes? There are two main competing channels to explain this:

\vspace{0.2cm}

\begin{itemize}  \setlength{\itemsep}{5pt}

    \item  \textbf{\emph{The binary channel (mass transfer):}} The binary channel was the original channel proposed to explain the process by which massive stars lose their outer layers to become cWR stars \citep{Paczynski1967}. The binary channel invokes a stellar companion that helps remove the outer layer of the primary via binary interaction. There are two main scenarios here. If the mass transfer is stable (thought to occur when the envelope of the donor star is not strongly convective and the mass ratio is not too extreme), then the companion accretes some or all of the mass of the primary, and the end result should be a binary with an inverted mass ratio: a WR star orbiting a more massive OB-type star. Many such systems are known, such as V444\,Cyg \citep{Marchenko1994, StLouis1993} in the MW, BAT99 19 in the LMC \citep{Shenar2019}, and SMC AB 8 in the SMC \citep{Shenar2016}. Alternatively, if mass transfer is not stable, common envelope (CE) evolution can ensue, leading to a dramatic decrease of the orbital period and an ejection of the outer layers of the primary star \citep{Iben1993, Ivanova2013}.  It is generally thought that the binary needs to avoid a merger if the envelope is to be ejected successfully, although recent empirical findings shed doubt on this assumption \citep{Shenar2023}. Post CE WR binaries were proposed (e.g., WR\,124; \citealt{Moffat1982, Toala2018}), albeit with ambiguous evidence. The WR X-ray binary Cyg\,X-3, thought to host a BH, appears to be a strong candidate for CE evolution \citep{vandenHeuvel1973, vanKerkwijk1992}.

    \item \textbf{\emph{The single channel (intrinsic mass loss):}} With the development of stellar-wind theory in the late 60s and early 70s of the past century, it became clear that massive stars can launch powerful stellar winds and lose significant amounts of mass throughout their main sequence and primarily post main sequence evolution stages such as the YSG and RSG phases \citep{Castor1975, deJager1988, Vink2001, Meynet2005, vanLoon2005, Vink2022}. To this come indications for variable, eruptive mass-loss which are thought to be associated objects such as LBVs \citep{Humphreys1994, Smith2011}, which some or all WR progenitors may have gone through. In the framework of the single-star channel, the WR progenitor thus ejects its outer layers via continuous mass loss or eruptive mass loss, without the help of a companion star. The single-star channel is also dubbed the "Conti scenario" after Peter Conti, who was one of the first to propose it.  The single-star channel is thought to occur above a certain metallicity-dependent mass threshold which marks the limit beyond which massive stars can self-strip themselves. This limit is highly uncertain and model-dependent. Evolution models typically predict self-stripping to occur above $M^{\rm strip}_{\rm ini} \approx 20 - 30\,M_\odot$ at solar metallicity, growing to $M^{\rm strip}_{\rm ini} \approx 35-100\,M_\odot$ at SMC metallicity (see table 1 in \citealt{Shenar2020}).  
\end{itemize}

\vspace{0.2cm}

The realization that massive stars are more commonly in binaries than not \citep{Kobulnicky2007, Sana2012, Sana2013, Moe2017}, and that these binaries will interact, gave a modern surge to the binary channel as a potentially dominant channel in forming cWR stars. Another argument against the single-star channel is the lower mass-loss rates predicted by modern mass-loss prescriptions for main sequence OB-type stars \citep{Krticka2017, Bjorklund2021} and for RSGs \citep{Beasor2020, Beasor2023}. 

It was originally thought that binaries should be increasingly important  for forming cWR stars at lower metallicities, since stellar winds  at low $Z$ are weak and binaries apparently offer the only viable way of stripping a star. However, as discussed in Sect.\,\ref{subsec:HeStars},  the WR phenomenon occurs for increasingly more massive stars as the metallicity drops, owing to the weaker winds of the stripped-star products.  These two effects -- the weaker winds of massive stars at low $Z$ on the one hand, but the higher-mass WR progenitors at low $Z$ on the other hand -- counteract each other. It is therefore not clear that binaries should indeed be more important for forming cWR stars at low $Z$ \cite[see also][]{Shenar2020}.

\begin{wrapfigure}{r}{0.5\textwidth} 
\vspace{-30pt}
\centering
\includegraphics[width=0.5\textwidth]{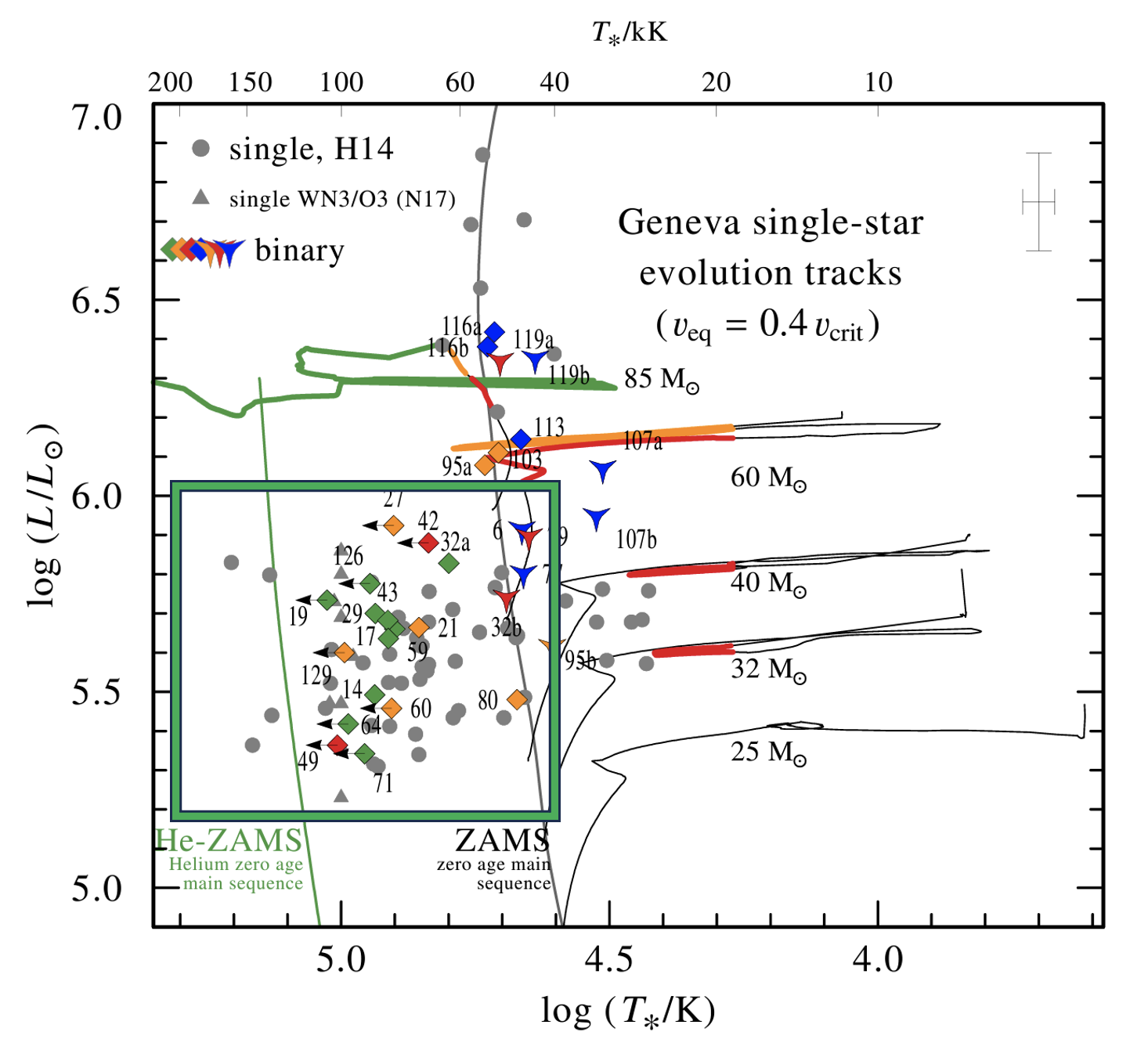}
\caption{HRD of the LMC WN population, showing the positions of apparently-single (gray symbols, \citealt{Hainich2015}) and binary (colored symbols, \citealt{Shenar2019}), compared to stellar evolution tracks computed with the Geneva evolution code for an LMC-like metallicity of $Z = 0.006$ \citep{Georgy2015}. The bulk of the apparently-single cWR population is evidently not reproduced by these tracks. Adapted from \citet{Shenar2019}. }  
    \label{fig:LMCWRs}
\end{wrapfigure}  
Evolution tracks, especially at low $Z$, struggle to efficiently remove the outer envelopes of massive stars via intrinsic mass loss (Fig.\,\ref{fig:HRDLMC}). However, observationally, it is not evident that binaries dominate the formation of cWR stars. A striking example for this is the population of 11 cWR stars in the SMC, which were monitored with low and high resolution spectroscopy (Sect.\,\ref{subsec:bin}), most recently by \citet{Schootemeijer2024}. Only 4/11 of these cWR stars have companions, and for the rest, companions can be effectively ruled out in the period and mass range that is thought to be relevant for binary mass-transfer to have played a role. A similar finding was recently reported for six low-luminosity WR stars in the LMC (dubbed WN3/O3 stars), which have too low initial mass to be stripped as single stars according to common evolution tracks, but which show no indications for companions in their vicinity \citep{Massey2023}.  

There are thus two options. Either 'standard' evolution models suffer from missing physics, or binary interactions occur in a way which is not predicted by 'standard' binary-evolution theory. The former option could have various solutions. The Humphreys-Davidson (H-D) limit (Fig.\,\ref{fig:HRDLMC}) implies that stars above $\approx 30\,M_\odot$ skip the RSG phase altogether, and suffer instead a short-lived eruptive phase. The H-D limit appears to be $Z$-independent and occurs at $\log L/L_\odot \approx 5.5$ \citep{Davies2018}, which coincides nicely with the onset of WR populations at LMC-SMC metallicities (Fig.\,\ref{fig:HRDLMC}). This implies that there could be a tight relation between the occurrence of the HD-limit and the formation of cWR stars; eruptive mass loss could be that missing link.  Eruptive mass-loss is typically not systematically implemented in evolution tracks, although efforts have been made to formalize its treatment  \citep{Owocki2004, SmithOwocki2006, Cheng2024}. Moreover, the internal mixing of massive stars is not well constrained in the literature, but may well be underestimated \citep{Higgins2019, Gilkis2021}; enhanced mixing makes the cores of the stars larger on expanse of their envelopes, and pushes them to a bluewards evolution. 

However, it is also possible that binary interactions result in seemingly-single WR stars. One possibility for this is that the currently single WR star was stripped by a companion which evolved more rapidly after mass transfer, and was ejected via a kick upon core collapse, disrupting the system altogether. In general, it is not expected that secondaries in massive binaries catch up with the evolution of the original primaries, even in case of conservative mass transfer, in which all mass is accreted onto the secondary \citep[e.g.][]{Eldridge2017, Marchant2023}. The main reason for this is the process of rejuvenation, which enriches the secondary with fresh hydrogen, thus resetting its evolution towards the zero-age main sequence. A way to circumvent this is by avoiding mixing of fresh hydrogen into the core of the secondary. Whether this occurs in nature or not is not clear. An alternative way to have binary interaction result in single stars is via stellar mergers. However, mergers are thought to either result in rapidly spinning stars \citep[e.g.][]{Vanbeveren2012, Gies2022} or strongly magnatized stars \citep{Ferrario2009, Schneider2019, Shenar2023, Frost2024}; it is not clear why a stellar merger should lead to the formation of a WR star a-priori.

While evidence for the dominance of binary interactions in forming WR stars is therefore ambiguous, two fact are noteworthy. First, the abundance of WR+O binaries in relatively short orbital periods (days to months), in which the O-type components exhibit substantial rotation \citep{Shenar2016, Shenar2019, Shara2017}, implies that binary mass transfer certainly took place in a significant number of systems; whether the WR star "ows" its formation to the companion (i.e., whether it would not have formed without it) is not clear. Secondary, it is interesting that while the close binary fraction (periods less than few years) of O-type is thought to be of the order of 70\% \citep{Sana2012}, the binary fraction of WR stars seems to be lowered. This suggests that at least some WR stars are the result of merging or other processes which led to the disruption of the binary, as discussed above.

\subsection{Final fates of cWR stars} \label{subsec:FinFate}

The fact that cWR stars are hydrogen depleted or free would seem to make them promising progenitors of type Ibc (i.e., hydrogen-free) core-collapse SNe \citep{Dessart2011, Langer2012, Yoon2017, Woosley2019}, though  empirical evidence for this is relatively sparse. Some SN events were suggested to originate from WN stars \citep[e.g.][]{Foley2007, Pastorello2008, Gal-Yam2014}, and, one recent SN observation, from a WC star \citep{Gal-Yam2022}. However, an increasing number of studies suggest that, at the upper-mass end, the cores of the stars are too compact to explode, and instead, the stars implode entirely \citep{Fryer1999, Sukhbold2016, Schneider2021}. Direct evidence for some direct collapse scenarios are implied from the orbits of binaries containing BHs such as Cyg\,X-1 \citep{Mirabel2003} and VFTS\,243 \citep{Shenar2022BH, VignaGomez2024}, which show no evidence for a past SN explosion. In general, it could be that some WR stars explode, while other directly implode, and that this could strongly depend on the final mass, rotation, and composition. The relative numbers of exploding vs.\ imploding WR stars are fully uncertain.

Similarly uncertain is the question regarding the compact object a WR star leaves behind upon collapse: neutron stars or BHs. In the Galaxy, WR stars may have masses as low as a few solar masses upon collapse, setting them in the neutron-star progenitor regime. At lower metallicities, where the WR phenomenon only  occurs at relatively high masses, and where WR stars generally lose less material, their final masses are typically larger ($\gtrsim 10\,M_\odot$), making them more likely to produce BHs \citep{Sukhbold2016}. However, the theory is still far from having a final mapping between the initial mass of a star and its final fate at core collapse, which requires sophisticated 3D modelling \citep[e.g.][]{Janka2012}.

\section{Concluding remarks} \label{sec:conclusions}

More than 150 years after their discovery, Wolf-Rayet stars continue to play a crucial role in our understanding of the upper-mass end, stellar feedback, dust formation, and stellar and binary evolution. While the underlying physical principle driver of their powerful winds -- stellar radiation and metallic line transitions -- are by now well established, the accurate modeling of their expanding amtospheres is still subject to approximations which are not reproduced by sophisticated hydrodynamically-consistent modelling. This hinders our ability to robustly derive the surface effective temperatures and stellar radii of WR stars. However, major advances to achieve such an accurate modelling have been reached in recent years that utilize both analytical, semi-analytical, and full blown computational power (Sect.\,\ref{sec:winds}), and 3D modelling has only just been initiated. This trend is likely to persist in the coming years.

We still cannot claim to know what the dominant formation channel of cWR stars is, and whether it is metallicity dependent. Taken at face value, observations suggest that the majority of cWR stars could form from their O-star progenitors without the need to invoke binary interactions. However, the multiple diverse routes binary evolution can advance makes it hard to rule out binary interactions as a dominant formation channel at this stage (Sect.\,\ref{subsec:bin}). 
Upcoming monitoring of WR populations in the LMC -- the largest resolvable population of WR stars at subsolar metallicity -- will take place with the 4MOST instrument between 2025 -- 2029 in the framework of the 1001MC survey \citep{Cioni2011}.  Monitoring of more distant extragalactic populations with next-generation telescopes such as the European Extremely Large Telescope (E-ELT) will take the field to lower metallicities, and will help boost the limiting statistics of current WR populations.

\seealso{\citet{Crowther2007} for a comprehensive review of WR stars}

\bibliographystyle{Harvard}
\bibliography{papers}

\end{document}